\documentclass[11pt,draftcls,onecolumn]{IEEEtran}
\usepackage[utf8]{inputenc} 
\usepackage[T1]{fontenc}    
\usepackage{hyperref}       
\usepackage{url}            
\usepackage{booktabs}       
\usepackage{amsfonts}       
\usepackage{nicefrac}       
\usepackage{microtype}      
\usepackage{lipsum}
\usepackage{fancyhdr}       
\usepackage{graphicx}       
\graphicspath{{media/}}     
\usepackage{amsmath,amsfonts}
\begin{document}
%
\title{Learning Rate-Compatible Linear Block Codes: An Auto-Encoder Based Approach}
%
%
%

\author{Yukun Cheng, Wei Chen,~\IEEEmembership{Senior Member,~IEEE}, Tianwei Hou,~\IEEEmembership{Member,~IEEE}, Geoffrey Ye Li,~\IEEEmembership{Fellow,~IEEE}, Bo Ai,~\IEEEmembership{Fellow,~IEEE}
\thanks{Yukun Cheng, Wei Chen, Tianwei Hou and Bo Ai are with School of Electronic and Information Engineering, Beijing Jiaotong University, China (Email:
ykcheng,weich,twhou,boai@bjtu.edu.cn). 

Geoffrey Ye Li is with Department of Electrical and Electronic Engineering, Imperial College London, UK (Email: geoffrey.li@imperial.ac.uk).}
}

\maketitle

\begin{abstract}
Artificial intelligence (AI) provides an alternative way to design channel coding with affordable complexity. However, most existing studies can only learn codes for a given size and rate, typically defined by a fixed network architecture and a set of parameters. The support of multiple code rates is essential for conserving bandwidth under varying channel conditions while it is costly to store multiple AI models or parameter sets. In this article, we propose an auto-encoder (AE) based rate-compatible linear block codes (RC-LBCs). The coding process associated with AI or non-AI decoders and multiple puncturing patterns is optimized in a data-driven manner. The superior performance of the proposed AI-based RC-LBC is demonstrated through our numerical experiments.
\end{abstract}


%
\IEEEpeerreviewmaketitle

\section{Introduction}
{H}igh-performance and flexible channel coding mechanism is a key enabler for the evolving key metrics of the forthcoming sixth-generation (6G) mobile communication system. As a vital class of channel codes, linear block codes (LBCs), including Bose-Chaudhuri-Hocquenghem (BCH) codes and Low-Density-Parity-Check (LDPC) codes, have demonstrated reasonable performance with adaptable complexity. 

To improve spectrum efficiency, the coding rate must adapt to changing channel conditions, therefore, the rate-compatible (RC) mechanism has been incorporated into LBC design \cite{5174525, 1347371, 6266764, 7045568}. Puncturing \cite{5174525, 1347371} is a direct approach to constructing RC linear block codes (RC-LBCs), which starts with a high-performance low-rate precode and systematically discards parity bits to increase coding rate.
The performance of puncturing-based codes is significantly influenced by two factors, i.e., precode properties and puncture patterns, typically optimized through coding theory\cite{richardson2008modern} and computer-assisted techniques \cite{8957447}. However, precodes are usually optimized for low rates, and code structures are designed without considering puncturing patterns, resulting in suboptimal performance \cite{6266764}.

RC-LDPC codes, as a representative of RC-LBCs, are adopted as a channel coding scheme for 5G NR \cite{3gpp.38.212}. Their performance is notably effective, particularly for long code lengths when employing decoders based on belief propagation (BP) or its variants. However, their high performance diminishes when dealing with shorter code lengths that have denser parity check matrices. 
Traditional code design procedures based on coding theory prioritize attributes such as Hamming distance\cite{richardson2008modern}. 
However, the desired performance may not always be attained when a decoder of low complexity is used to facilitate practical application\cite{8890904}. For example, LBCs designed for minimum distance guarantee are only optimal under conditions of maximum likelihood (ML) decoding.

Artificial intelligence (AI) offers additional benefits when conventional techniques reach their limits or specific design criteria are difficult to meet using traditional approaches, especially in shorter-lengths.
These studies can broadly be categorized into AI-based decoders and AI-based code design. A popular class of AI-based decoders is unfolding iterative algorithms into neural networks \cite{7852251, 9913203, 9427170}. For AI-based code design, neural networks are utilized for non-linear encoding \cite{9839215, 9438648}. Additionally, reinforcement learning \cite{8890904} and self-supervised learning \cite{9896912} based methods are proposed to generate structured error-correcting codes. However, existing AI-based code neglects the importance of adaptable code rates \cite{1146084}. Both the transmitter and receiver need to retain multiple AI models for different code rates. Meanwhile, current AE-based code design constrains the parity-check matrix structure to a systematic form \cite{9896912}. This may lead to more short cycles, adversely affecting the performance of the BP-based decoders \cite{7852251}. 

In this article, we present an AI-based RC-LBC capable of accommodating various code rate requirements.
We leverage concepts from traditional puncturing-based RC codes and propose a modified AE with learning-based RC-LBCs. 
Rather than focusing on the decoding performance of BP-based neural decoders in the AE, we introduce an AI-based technique for designing RC-LBCs that work efficiently with the specified BP-based decoders and the given puncture pattern, where we adopt a simple truncation of the end bits in this work. Optimizing the puncturing patterns may also improve the performance of this coding scheme, which is left for future work. These learned codes are interpretable and applicable to existing communication systems.
It is an innovative effort to enable AI-based RC codes. Furthermore, the designed AE can optimize parity-check matrices with systematic and some non-systematic structures and is compatible with AI or non-AI decoders. 

\section{Preliminary}
In this section, we briefly introduce the concept of RC-LBCs, rate-compatible neural belief propagation (RC-NBP) decoder, and multi-task learning (MTL).

\subsection{System Model}
\begin{figure}[!t]
    \centering    \includegraphics{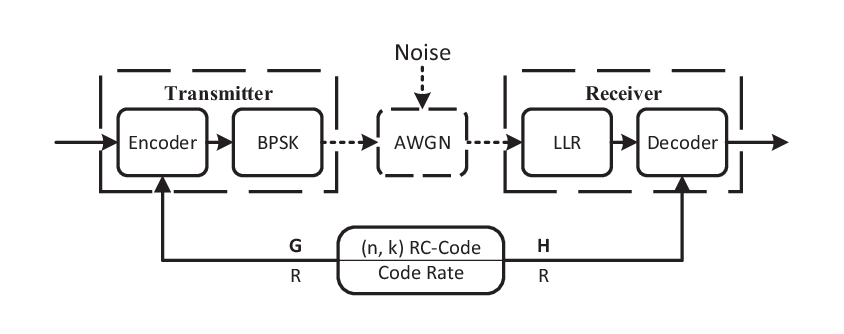}
    \caption{The system model with RC code.}
    \label{Sysmodel}
\end{figure}

Akin to \cite{9896912}, we consider the communication system shown in \figurename \ref{Sysmodel}. Information bits are processed at the transmitter, which includes a channel encoder and a modulation module of binary phase shift key (BPSK). The encoded symbols are transmitted to the receiver through an additive white Gaussian noise (AWGN) channel, where the noisy symbols are converted to LLRs and sent into a channel decoder. Generate matrix $\mathbf G$ and parity-check matrix $\mathbf H$ of code $C(k,n)$  guide the encoding process at the encoder and establish the neuron connection pattern within the decoder, respectively. Code rate $R$ dictates the puncturing procedure in the encoder and neuron activation status in the decoder.

\subsection{Rate-Compatible Linear Block Codes}
In this article, we denote a binary LBC of length $n$ and information bits of length $k$ as $C(n,k)$, and the code rate is defined as $R=k/n$. Let $\mathbb{F}_2$ denotes the binary finite field, $\mathbf{G} \in \mathbb{F}_2^{k, n}$ and $\mathbf{H} \in \mathbb{F}_2^{(n-k),n}$ be the corresponding generate matrix and parity-check matrix, respectively. 

RC-LBCs, different from conventional LBCs, have a nested structure among codewords of different rates. This allows the code rate to be adjusted flexibly through puncturing the redundant bits.
Assume that two codewords $\mathbf{c}_1$ and $\mathbf{c}_2$ are derived from the same precode $C_0(n,k)$ with different rates $R_1$, and $R_2$ $(R_1 > R_2)$, respectively, and $\mathbf{c}_1$ is included within $\mathbf{c}_2$. 
In this work, we adjust the code rate by discarding parity bits, while keeping $k$ constant throughout the process, and lowest-rate code is denoted as precode.

\subsection{Rate-Compatible Neural Belief Propagation Decoder}
\begin{figure}[!t]
    \centering
    \includegraphics{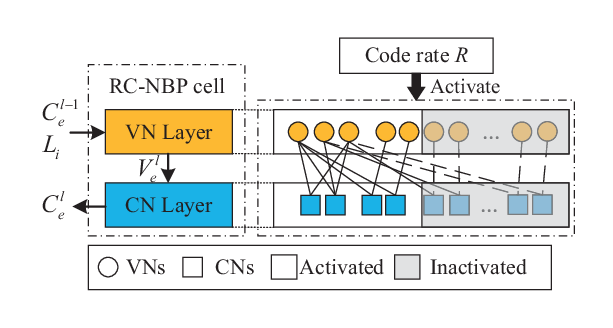}
    \caption{Structure of RC decoding cell}
    \label{Decoder}
\end{figure}

The Neural Belief Propagation (NBP) decoder is built by unfolding the conventional BP algorithm into a neural network. Learnable weights/biases are introduced to adjust the messages transferred among neurons, which effectively mitigate the drawbacks brought by short cycles. In this article, we adapt the RC-NBP decoder proposed from our previous study \cite{10365399} while each decoding iteration is reformed to an RC-NBP cell depicted in Fig. \ref{Decoder}. The decoder dynamically adjusts the activation status of neurons according to code rate, conserving computing and storage resources, tries to learn learns optimal weights for multiple rates. 

Now, we briefly introduce the decoding process. Define $\mathbf{H}$ as a Tanner Graph of variable nodes (VNs) and check nodes (CNs) connected by edges $e$. RC-NBP unfolds the iterative BP algorithm into a neural network, encapsulating every iteration and edge into a RC-NBP cell and a neuron, respectively. 
For a received noisy codeword $\Tilde{\mathbf{c}}$, which corresponds to codeword $\mathbf{c}$, the Log-Likelihood-Ratio (LLR) value of the $i$-th bit $\Tilde{\mathbf{c}}_i$ is defined as $L_i = \ln (P(\Tilde{\mathbf{c}}_i|\mathbf{c}_i=0)/P(\Tilde{\mathbf{c}}_i|\mathbf{c}_i=1)),$
where $P(\hat{\mathbf{c}}|\mathbf{c})$ denotes the transition probability from $\mathbf{c}$ to $\Tilde{\mathbf{c}}$. Let $e=(i,j)$ represent the edge connecting the $i$-th VN and the $j$-th CN. 


Let $l$ denote the number of the RC-NBP cell corresponds to the l-th iteration of the decoder.
For the $l$-th RC-NBP cell, the CN to VN (C2V) message and the VN to CN (V2C) message are defined as $C_e^l$ and $V_e^l$, respectively. 
The V2C messages, $V_{e}^{l}$, is calculated using
\begin{equation}
	\label{BPVC}
	V_{e}^{l} = L_i + \sum_{e'\in\varepsilon_i \backslash j}C_{e'}^{l-1},
\end{equation}
where $\varepsilon_i$ denotes the set of edges adjacent to the $i$-th VN, and $\varepsilon_i \backslash j$ denotes the set excluding the edge adjacent to the $j$-th CN. At the first RC-NBP cell, $C_e^0$ is initialized as 0.
The C2V message, $C_e^l$, is calculated using
\begin{equation}
\label{NNBPCV}
C_e^l = \alpha_e^l\times2\textnormal{tanh}^{-1}\left(\prod_{e'\in\varepsilon_j \backslash i}\textnormal{tanh}\left(\frac{V_{e'}^{l}}{2}\right) \right)+\beta_e^l,
\end{equation}
where $\alpha_e$ and $\beta_e$ are learnable weight and bias associated with the edge $e$, respectively, $\varepsilon_c$ denotes the set of edges adjacent to the $j$-th CN, and $\varepsilon_j \backslash i$ denotes the set excluding the edge adjacent to the $i$-th VN. Both multiplications and biases affect the messages passed through the edges and appropriate settings bring faster convergence and better BER performance \cite{9427170}. 

For the output layer, the decoded LLR value of $\mathbf{c}_i$ is calculated by $O_i = L_i + \sum_{e\in \varepsilon_v}C_{e}^{l_{max}},$
where $l_{max}$ denotes the maximum number of RC-NBP cells. Then the $i$-th decoded bit $\hat{\mathbf{c}}_i$ can then be obtained using $\hat{\mathbf{c}}_i = [1-sgn(O_i)]/2$. When the code rate changes, the codeword is encoded with same $\mathbf G$ but punctured before transmission, and only portion of $\mathbf H$ is utilized for decoding higher-rate codes. Consequently, only a subset of the VNs and CNs participate in the decoding process, presented as some neurons are turned on or off in the RC-NBP decoder.
This approach allows the transmitter and decoder to support multiple code rates with a single parameter set.
For more details of the decoder, interested readers are referred to \cite{10365399}.

\subsection{Multi-Task Learning}
MTL aims to train a model that fits multiple related tasks. Models trained on datasets from multiple tasks have a lower risk of overfitting and typically yield better performance \cite{9392366}. 
The main goal here is to train a code capable of supporting the requirement for multiple code rates, which matches the core idea of MTL. 
We regard the communication task under different coding rates as distinct tasks, requiring the proposed model to effectively learn an RC-LBC under the considered conditions. We devise a parameter-sharing scheme to exploit the nested structure of RC codes, where the parameters utilized for higher-rate codes are also trained during the training process for lower-rate codes.

\section{Proposed Rate Compatible Linear Block Code Auto-encoder}
We focus on the design of AE for channel coding utilizing RC-LBCs. The employed codes are optimized through a data-driven approach, differing from the traditional optimization approach based on coding theory. We employ an RC-NBP decoder and a predetermined puncture pattern. It is anticipated that the learned codes will exhibit satisfactory performance under these constraints.

\subsection{Proposed Auto-Encoder Architecture}
As shown in Fig. \ref{Enc}, the proposed AE consists of a matrix generating (Matrix-Gen) module, and a channel encoder module, an RC-NBP decoder. The learnable parameters include two parts, i.e., coding parameters $W$ used to generate parity-check matrix $\mathbf{H}$, and the decoding parameters introduced in subsection II-B. The modules applied in our AE are detailed in the following subsections. We will now provide a brief overview of the computational process of the AE.

\subsubsection{Initialization}
Assume we have an RC-LBC, with the lowest rate $R_0=k/n_0 < R$, where $R$ denotes the desired code rate under the current channel condition. The Matrix-Gen module generates $\mathbf{H}$ and $\mathbf{G}$ based on coding parameters $W$. $R$ dictates both the puncture process in the encoder and the neuron activation status in the decoder. During the training stage, $\mathbf{H}$ is fed into both the encoder and decoder, while in the inference stage, $\mathbf{G}$ and $\mathbf{H}$ are separately fed into the encoder and decoder.

\subsubsection{Encode}
A $k$-length information bit vector, denoted as $\mathbf{x}$, is encoded by multiplying it with $\mathbf{G}$. 
The encoding process using $\mathbf{G}$ can be written as $\mathbf{c}=\mathbf{x}\mathbf{G}$, with the coded bits $\mathbf{c}$ satisfying $\mathbf{c}\mathbf{H}^T=\mathbf{0}\in\mathbb{F}_2^{n-k}$. 
$\mathbf{G}$ are generated by the Matrix-Gen module. The encoded bits are then punctured, starting from the last redundant bits, into a codeword $\mathbf{c}$ according to $R$, with a length of $n_c$. $\mathbf{c}$ is then modulated with BPSK and transmitted over an AWGN channel.

\subsubsection{Decode}
Upon reception, the LLRs of the received symbols are computed and inputted into the decoder. The connection patterns of the decoder neurons are determined by $\mathbf{H}$ \cite{10365399}. After decoding, the LLR values of recovered information bits $\hat{\mathbf{x}}$ are extracted as the first $k$ LLR values from the decoded results. Subsequently, a sigmoid function is employed to assign them binary-like values.

\begin{figure}[!t]
	\centerline{\includegraphics{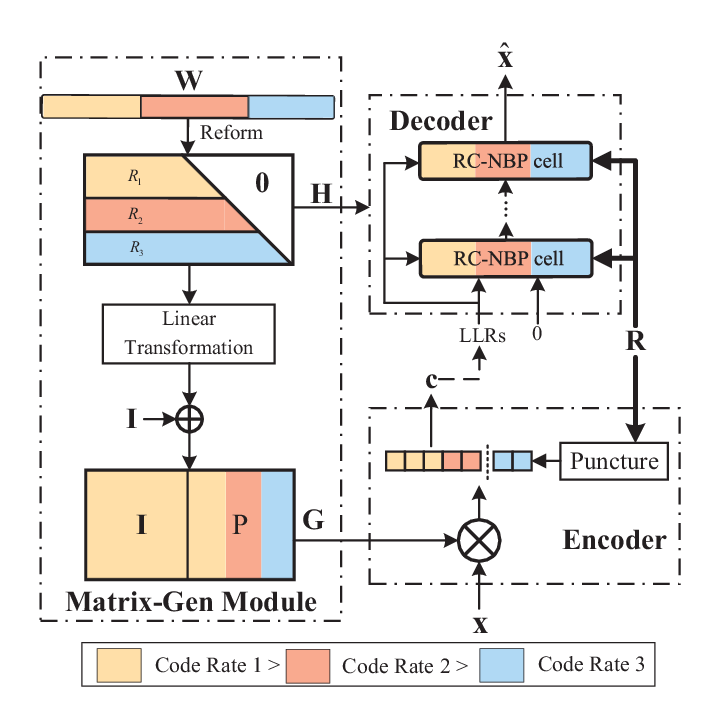}}
	\caption{The architecture of proposed auto-encoder.}
	\label{Enc}
\end{figure}

\subsection{Encoder and Matrix-Gen Module}
In the initialization process, the Matrix-Gen module derives $\mathbf{H}$ and $\mathbf{G}$ from coding parameters $W$, which are initialized randomly within the range $[-0.01, 0.01]$. In the encoding process, $\mathbf{G}$ is then utilized to encode information bits.
A direct encoding method is multiplying the information bits by $\mathbf{G}$, derived from $\mathbf{H}$ in $\mathbb{F}_2$. 

However, the non-deterministic number of operations in the Gauss-Jordan elimination makes it difficult to apply automatic differentiation with deep learning tools, since the computational graph needs to be updated for each change in $\mathbf{H}$.
To circumvent this challenge, we added some restriction to the structure to obtain a fixed computation graph during training. The considered $\mathbf{H}$ is of the form $\mathbf{H} = [\mathbf{H}_1|\mathbf{H}_2],$
where $\mathbf{H}_1$ is an $(n-k)\times k$ matrix and $\mathbf{H}_2$ is a lower triangular matrix of size $(n-k)\times(n-k)$. The form of $\mathbf{H}_2$ is
\begin{footnotesize} 
\begin{equation}
	\mathbf{H}_2=
	\begin{bmatrix}	
		1\\
		h_{2,k+1}  & 1\\
		h_{3,k+1} & h_{3,k+2}& 1 & \multicolumn{2}{c}{\raisebox{1.3ex}[0pt]{\Huge0}}\\	
		\vdots & \vdots& \ddots &\ddots \\	
		
		h_{m,k+1}& h_{m,k+2} &\cdots & h_{m,n-1} &1
	\end{bmatrix},
\end{equation}
\end{footnotesize}
where the diagonal elements and the upper triangular part of $\mathbf{H}_2$ are set to be 1 and 0, respectively. And all remaining elements of $\mathbf{H}$ can be optimized during training.

To ensure the encoding matrix generated from float-type parameters is both binary and differentiable, we applied a differentiable step function (DSF) as proposed in \cite{9896912}. In the forward pass during the training stage, a non-differentiable step function maps the negative/positive values of $W$ to 0/1, respectively. In the backward pass, a sigmoid function is empolyed as a differentiable approximation of the step function, which can be written as: 
\begin{equation}
	\frac{df(x)}{dx} \approx \frac{d\sigma(x)}{dx}=\sigma(x)\sigma(1-x),
\end{equation}

where $f(\cdot), \sigma(\cdot)$ denotes the non-differentiable step function and sigmoid function, respectively. 

As shown in Fig. \ref{Enc}, The mapped elements constitute the learnable part of $\mathbf{H}$ while an identity matrix is added to form the diagonal elements in $\mathbf{H}_2$.
To obtain $\mathbf{G}$, we can use linear transformations in $\mathbb{F}_2$ to reform $\mathbf{H}$ into a systematic structure, denoted as $\mathbf{H} = [\mathbf{P}|\mathbf{I}]$, where $\mathbf{I}$ and $ \mathbf{P}$ denote an identity matrix and a redundancy matrix, respectively. Then we have $\mathbf{G} = [\mathbf{I}|\mathbf{P}^T]$. Information bits are encoded through matrix multiplication followed by a mod-2 calculation, with the deep learning platform providing tools for approximating gradients.

\subsection{Rate-Compatible Decoder}
We incorporate the RC-NBP decoder \cite{10365399}, as described in Section II-B, into our AE framework with a major modification.
Previous studies \cite{10365399, 9427170} have introduced learnable decoding parameters for edges corresponding to ``$1$"s in $\mathbf{H}$. In this AE, we allocate a decoding parameter for every potential edge in each cell\cite{9896912}, although only a subset of these parameters are utilized during decoding. 
For expedited training and reduced parameter number, we only adapt weight multipliers in our experiments.

\subsection{Code Settings and Training Procedure}
In this section, we provide an overview of the settings, the training procedure.

\subsubsection{Puncture Pattern}
We begin our design with a precode at the lowest rate, and then gradually discard the last parity bit to increase the code rate. 
The proposed AE architecture anticipates that the learned codes will be optimized given the existing puncturing pattern.

\subsubsection{Training Procedure}
Before learning an RC-LBC, we need to acquire a high-performance precode at a low rate. During this process, the decoder weights are set to $1$, mirroring a conventional BP decoder. Once a precode has been acquired, the training proceeds with a set of randomly generated codewords spanning a variety of rates.

We use binary cross-entropy (BCE) between the decoded bits $\hat{x}$ and information bits $\mathbf{\mathbf{x}}$ as the loss function, which is written as
\begin{equation}
	Loss(\hat{\mathbf{x}},\mathbf{x})=-\frac{1}{k}\sum_{i=1}^{k}\mathbf{x}_ilog(\hat{\mathbf{x}}_i)+(1-\mathbf{x}_i)log(1-\hat{\mathbf{x}}_i),
\end{equation}

where $\mathbf{x}_i$ denotes the $i$-th bits of the word $\mathbf{x}$. By exploiting MTL, the parameters associated with matrices of multiple code rates are trained using a variety of datasets. Specifically, parameters associated with a specific code rate would be learned from all data with code rates no higher than that rate, e.g. the optimization step of $W_{i}$ can be written as $W_{i} = W_{i}-\sum_{r} \frac{\partial Loss_r}{\partial W_{i}}$, where $Loss_r$ denotes the BCE loss of the $r$-th highest code rate.

\section{Evaluation Results}
In this section, we construct the AE with learned RC-LBCs and evaluate their bit-error rate (BER) performance. The complexity of encoding is linearly related to the codeword length, with the main complexity arising from the decoder. For a detailed discussion of the complexity, reader can refer to \cite{10365399}.
The proposed AE can optimize matrices with both systematic and non-systematic structures and can be associated with both AI and plain BP decoders.
All decoders are set to execute efficiently 5 iterations for a fair comparison.
The RC-LBCs are trained using the Adam optimizer with a learning rate of $0.001$. The batch size is 256 and each epoch contains 2048 randomly generated vectors, where the codewords in each batch are encoded based on a same matrix. The training process consists of 5,000 epochs with a single code rate for training precode and 5,000 epochs with mixed code rates for training AI-RC-LBC, with even batches for each code rate. The code rate information is predefined and stored in the proposed AE before training.

The $``\emph{Curse of Dimensionality}"$ constrains the length of AI-based channel codes, where the number of codewords associated with information blocks of size $k$ is $2^k$. As the code length increases, optimizing AI-based codes in a data-driven manner becomes increasingly challenging. Consequently, AI-based codes encounter difficulties in achieving comparable performance with larger code lengths \cite{8054694, 9896912}. Addressing this issue remains a topic for future research, which lies beyond the primary focus of this work.

\begin{figure}[!t]
	\centerline{\includegraphics{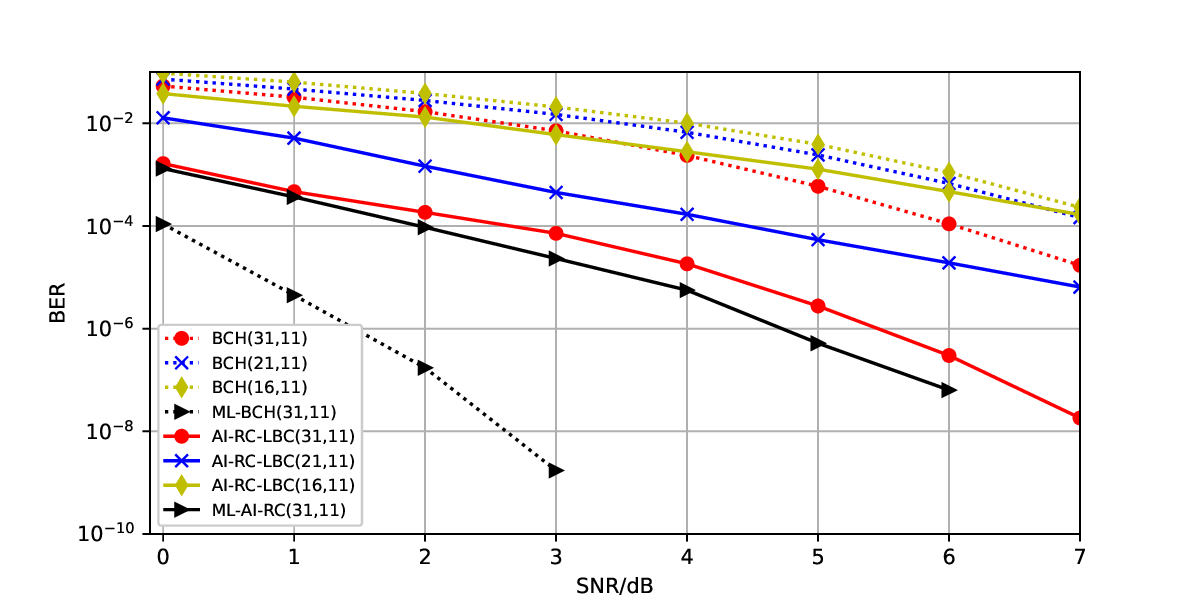}}
	\caption{AI-RC-LBC (k=11) vs. BCH codes under BP decoder.}
	\label{RCvsBCH}
\end{figure}

\begin{figure}[!t]
	\centerline{\includegraphics{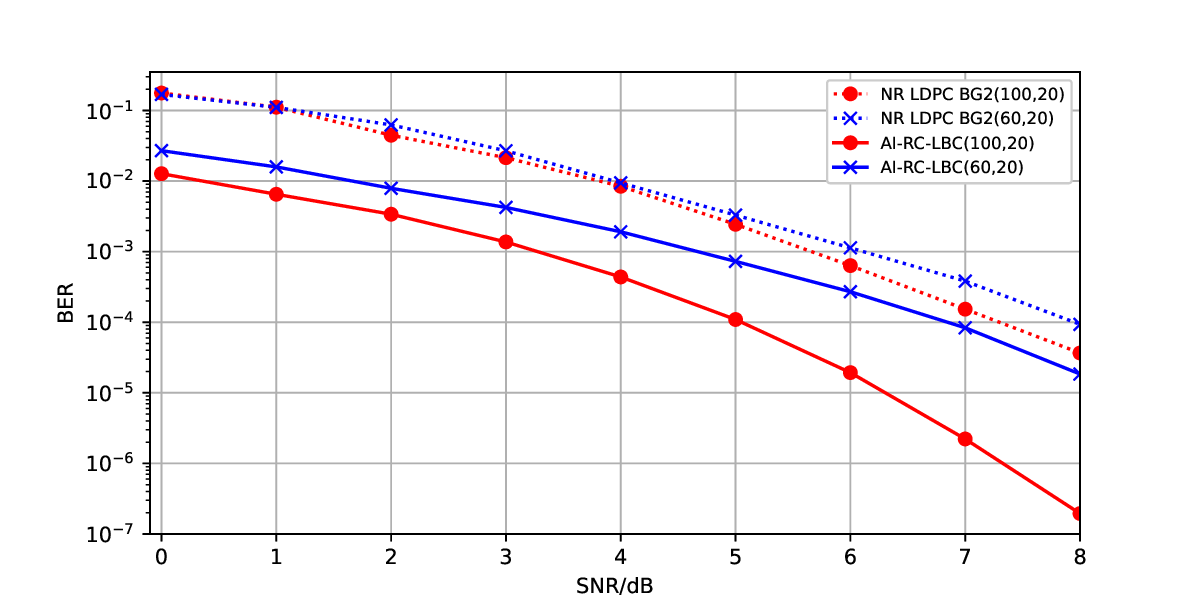}}
	\caption{AI-RC-LBC (k=20) vs. NR LDPC BG2 codes under BP decoder.}
	\label{RCvsLDPC}
\end{figure}

\subsection{AI-RC-LBCs vs. BCH \& LDPC codes}

We validate the proposed AE using AI-designed RC-LBCs, denoted as AI-RC-LBCs, by conducting experiments with a systematic matrix structure and conventional BP decoding algorithm. 
We choose $(31,11)$, $(31,16)$, and $(31,21)$ BCH codes and $(100,20)$, $(60,20)$ BG2 code defined in 5G standard\cite{3gpp.38.212} as baselines.
For comparison, we train two AI-RC-LBCs with lowest rates of $11/31$ and $20/100$, respectively, following the baseline codes. We obtain $C(21,11)$, $C(16,11)$\footnote{For the same SNR, the longer the code length, the lower the code rate, the better the BER performance. Thus, if $C(21,11)$ and $C(16,11)$ by puncturing perform better than (31,16) and (21,31) BCH codes, respectively, we can infer that the proposed approach has improvement.}, and $C(60,20)$ by puncturing. The training SNRs for different rates in AI-RC-LBC with $k=11$ and $k=20$ are set at [3, 4, 5] dB and [4, 5] dB, respectively, where the corresponding baseline codes achieve a BER less than $10^{-2}$.

As shown in Fig. \ref{RCvsBCH} and Fig. \ref{RCvsLDPC}, the learned AI-RC-LBC outperforms BCH codes and BG2 LDPC codes in all cases with similar code rates. At the lowest rate and a BER of $10^{-4}$, the margin is 3dB and 2dB compared to BCH and LDPC codes, respectively.
As the code rate increases, the performance gap between the AI-RC-LBC and baseline codes decreases, which is the result of the shorter code lengths of the punctured codewords. From the figures, the learned RC-LBC can replace multiple single-rate conventional codes, which saves storage resources and supports rate-matching schemes with improved BER performance.

\subsection{AI-RC-LBC vs. AI-based LBC}
\begin{figure}[!t]
	\centerline{\includegraphics{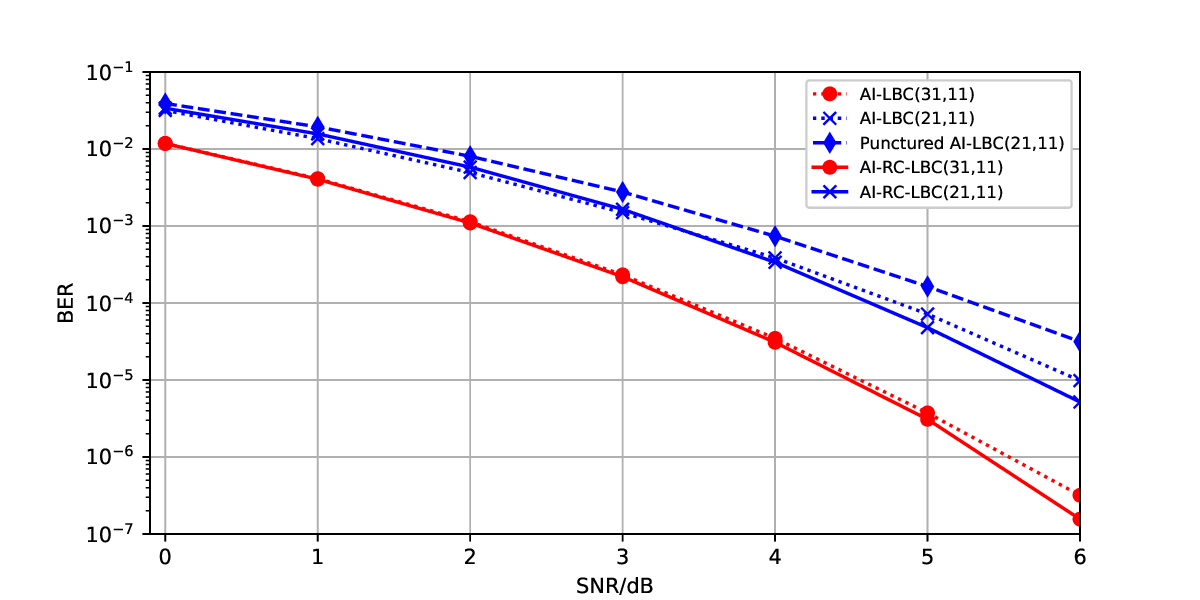}}
	\caption{AI-RC-LBC vs. AI-LBCs under neural decoders.}
	\label{RCvsSingle}
\end{figure}

To investigate how training with multiple code rates affects the learning of LBCs, we compare AI-RC-LBC with single-rate AI-based codes, denoted as AI-LBCs here. The systematic matrix structure and neural BP decoders are adopted in this experiment.
The AI-RC-LBC is derived from a precode of $C(31,11)$, and code $C(21,11)$ is obtained through puncturing. We conduct comparison with AI-LBCs $C(31,11), C(21,11)$. For a fair comparison, AI-LBCs are trained with 10,000 epochs with a single code rate, which is of the same training codeword numbers as AI-RC-LBCs.
The training samples of two considered code rates are generated at SNRs of 3dB and 4dB, respectively. We apply the puncturing pattern used in RC-LBC to the (31,11) AI-LBC, which generates punctured AI-LBC $C(21,11)$. As shown in Fig.~\ref{RCvsSingle}, the RC-LBCs outperform AI-LBCs at the same rates, particularly in high SNR regions. This suggests that the model benefits from a more diverse dataset. Meanwhile, the diverse scenarios act as a form of regularization, which is also a key advantage of MTL.
Furthermore, among the (21,11) codes, the punctured AI-LBC performs the worst, further confirming that the MTL guides the AE to learn a matrix suitable for variable code rates.

\begin{figure}[!t]
	\centerline{\includegraphics{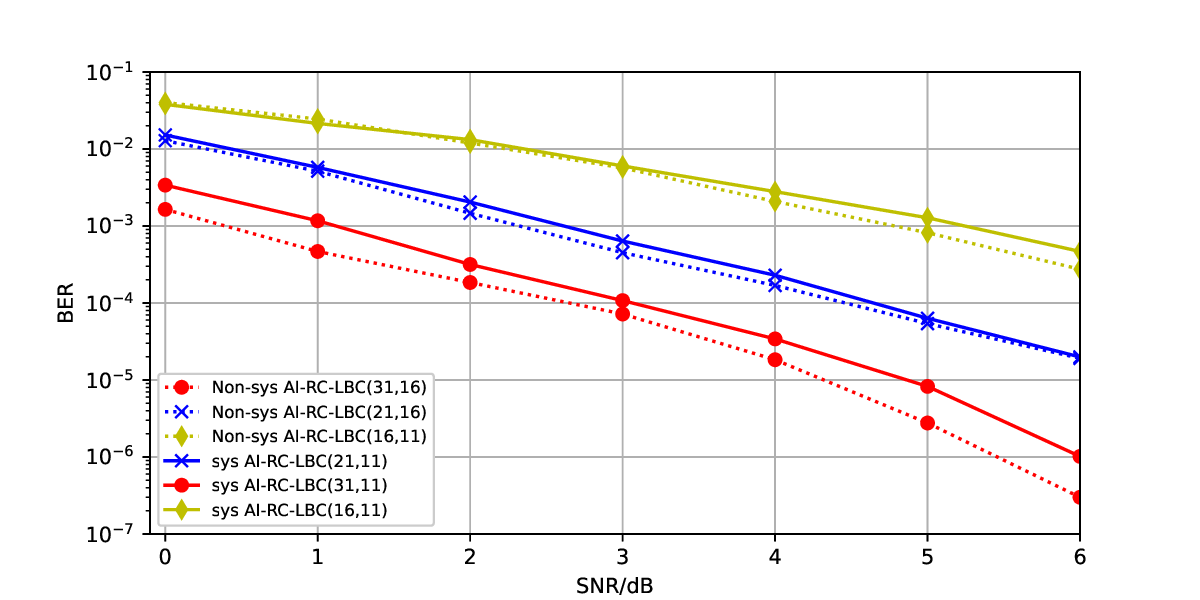}}
	\caption{Non-systematic RC-LBCs vs. systematic RC-LBCs.}
	\label{Sys}
\end{figure}

\subsection{Non-systematic structure vs. Systematic structure}
Our encoding scheme extends the learnable matrix $\mathbf{H}$ to some non-systematic form, which is expected to yield improved performance when used with BP-based decoders. In this subsection, we compare the performance of AI-RC-LBCs that are trained with non-systematic $\mathbf{H}$ matrices to those that are trained with systematic ones, neural decoders are considered in this experiment. 
As shown in Fig. \ref{Sys}, non-systematic codes, $C(31,11)$, $C(21,11)$, and $C(16,11)$, outperform systematic codewords. In conclusion, the considered structure achieves better performance with BP-based decoders. The margin is increased when the code length is greater or the decoder runs fewer iterations, resulting in a sparser matrix with fewer short cycles or a decoder more affected by short cycles.

\section{Conclusion}
In this article, we present a novel AE architecture for an AI-based rate-compatible channel coding scheme. The deployed channel codes, RC-LBCs are automatically learned according to a given decoder and puncturing pattern.
It is demonstrated that AI-RC-LBC outperforms the ones deployed BCH codes and AI-based single-rate LBCs under short code block lengths and multiple code rates.
The acquired AI-RC-LBC can substitute multiple single-rate codes to save storage. Moreover, the proposed AE can also learn some non-systematic LBC matrices.


%

\appendices
\section{Proof of the First Zonklar Equation}
Appendix one text goes here.

\section{}
Appendix two text goes here.

\section*{Acknowledgment}

The authors would like to thank...

\ifCLASSOPTIONcaptionsoff
  \newpage
\fi

\bibliographystyle{IEEEtran}

\bibliography{reference.bib}

\end{document}